\documentclass[prd,showpacs,amsmath,amssymb,superscriptaddress,preprintnumbers,nofootinbib]{revtex4}

\usepackage{amssymb}
\usepackage{amsmath}
\usepackage{graphicx}

\begin{document}

\newcommand{\Fst}{{\mathop {\rule{0pt}{0pt}{F}}\limits^{\;*}}\rule{0pt}{0pt}}

\title{Nonminimal dyons with regular gravitational, electric and axion fields}

\author{Alexander B. Balakin}
\email{Alexander.Balakin@kpfu.ru} \affiliation{Department of
General Relativity and Gravitation, Institute of Physics, Kazan
Federal University, Kremlevskaya str. 16a, Kazan 420008, Russia}

\author{Dmitry E. Groshev}
\email{groshevdmitri@mail.ru} \affiliation{Department of
General Relativity and Gravitation, Institute of Physics, Kazan
Federal University, Kremlevskaya str. 16a, Kazan 420008, Russia}

\begin{abstract}
We consider the distribution of the axionic dark matter and the profiles of the axionically induced electric field on the background of a regular magnetic monopole.
The background static spherically symmetric spacetime is considered to be regular at the center of the object and has no event horizons, the nonminimal interaction of strong gravitational and monopole type magnetic fields being the source, which provides this regularity. We assume that the pseudoscalar (axion) field also is nonminimally coupled to the electromagnetic and gravitational fields, and we search for solutions to the extended equations of nonminimal axion electrostatics, which satisfy two requirements: first, the axionically induced radial electric field vanishes at the center; second, the  pseudoscalar field is regular at the center of the formed axionic dyon. We have found the set of parameters, describing the nonminimal axion-photon coupling, for which both requirements are satisfied. Properties of the two-parameter family of regular solutions to the equations of nonminimal axion electrostatics are analyzed numerically. The regular axionically induced electric field is shown to have the structure, which is typical for a simple electric layer.
\end{abstract}

\pacs{04.20.Jb, 04.40.Nr, 14.80.Hv}

\maketitle

\section{Introduction}

The term {\it axionic dyon} entered the scientific lexicon after publication of the paper \cite{Wilczek2}, in which Wilczek  considered a magnetic monopole surrounded by the axion shell as a source of the induced electric field, and as a reason of appearance of an effective electric charge, which could be detected by a distant observer. Now the model of the axionic dyon  is used widely for the theoretical investigation of the structure of the objects with strong gravitational, magnetic, electric and pseudoscalar (axion) fields. The interest to such field configurations is motivated, in particular, by the problem of identification of the particles, which form the cosmic dark matter. There is the well documented story, how  Peccei and Quinn have predicted \cite{a1} and Weinberg \cite{a2} and Wilczek \cite{a3} have introduced  into the High Energy Physics a new important element, the light massive pseudo-Goldstone boson, indicated later as the axion. Now the axions are considered as the constituents of the cosmic dark matter (see, e.g., \cite{a4,a5} for details and references). The massive dark matter axions are assumed to be concentrated near the objects with strong gravitational and magnetic field, thus, the non-uniformity of its distribution switches on the mechanism of the electric field induction. This mechanism is predicted in the framework of the axion electrodynamics, which was elaborated by Ni \cite{Ni} and Sikivie \cite{Sikivie}, and has a lot of applications in astrophysics and cosmology.

In order to use the axionic dyons as a theoretical laboratory for investigation of the dark matter particles, one needs to know fine details of the dyon structure.
In this context, Lee and Weinberg studied in \cite{LW91} the problem of horizons and thus opened the discussion concerning the causal structure of the axionic dyons. But when one studies the causal structure of spherically symmetric objects, the additional question arouses interest: what is the behavior of the gravitational, electromagnetic and axion fields near the center? This question is associated with a search for the solutions to the model master equations, which are regular at the center. The problem of absence of the central spacetime singularity in the spherically symmetric objects has been raised by Bardeen in \cite{Bardeen}, and during next fifty years this problem has been studied in many aspects (see, e.g., \cite{r1,r2,r3,r4,r5,r6,r7,r8,r9,r10,r11,r12,r13,r14,r15,r16,r17}); mostly, the {\it nonlinear} electromagnetic field was considered as the guarantor of the regularity. {\it Nonminimal} coupling of the gauge and electromagnetic fields to the spacetime curvature, as a cause of regularity, was discussed in the papers \cite{R1,R2,R3,R4,R5,R6}. In other words, there are at least two theories: first, the nonlinear electrodynamics, second, the nonminimal field theory, in the framework of which one can provide the regularity of the gravitational field at the center of the magnetic monopoles.

However, one has to keep in mind the following specific detail of the obtained exact solutions. For the models of the nonminimal Wu-Yang and Dirac monopoles the spacetime is regular, i.e., the metric function, the Ricci scalar, and all the quadratic curvature invariants are finite, while the scalar square of the magnetic field is infinite at the center \cite{R1,R2,R3,R4,R5,R6}. On the contrary, there are nonminimal models with the electric field regular at the center (see, e.g., \cite{BBL,BZ}), however, the gravity field has the mild singularity, i.e., the metric is regular, but the Ricci scalar takes infinite value at the center. In other words, exact solutions to the nonminimal field equations, for which both gravitational and electromagnetic fields are regular at the center are not yet presented.

Let us imagine now that the pseudo-scalar (axion) field is added as a new ingredient to the set of interacting regular gravitational and singular magnetic fields. This axion field is considered to be nonminimally coupled with both: the gravitational and electromagnetic fields (see \cite{BWTN2010} for the details of the corresponding model). We are interested to answer the following question: whether the regular at the center solutions for the electric and axion fields exist? In other words, behavior of which field, gravitational or magnetic, the axiono-electric field inherits? Below we show that a specific choice of coupling parameters in the nonminimal Einstein-Maxwell-axion theory allows us to construct the model, which admits regular solutions for the pseudoscalar and axionically induced electric fields surrounding the magnetic monopole with regular gravitational field.

The paper is organized as follows. In Section II we consider the mathematical formalism of the presented nonminimal model; in Subsection IIA we recall the exact solutions describing the regular background gravitational field formed by the strong monopole type magnetic field; in Subsection IIB we extend the formalism of the nonminimal model by including the electric and axion fields; in Subsection IIC we reduce the master equations of nonminimal axion magnetoelectrostatics using the ansatz about the spherical symmetry. In Section III we study the solutions with regular at the center electric and axion fields; in Subsection IIIA we represent the specific set of nonminimal coupling constants, for which this regularity is guaranteed; in Subsection IIIB we present the results of  numerical analysis of the regular model. Section IV contains discussion of the obtained results.

\section{The formalism}

\subsection{The background gravity field}

\subsubsection{On the exact background solution for the nonminimal model}

In this work we follow the hierarchical approach and use the two-level nonminimal model. The basic level relates to the background gravitational field formed by the baryon matter and strong magnetic field of the monopole type nonminimally coupled to the spacetime curvature. The background problem is already solved, the metric regular in the center is found in our previous papers (see, e.g., \cite{R1,R2,R3,R4,R5,R6}). Nevertheless, below we recall main details of this story. The second level of the model is connected with the interacting axion and electromagnetic fields in the given spacetime background; this part of the work is the new one. The total action functional is presented, respectively, by the following sum:
\begin{equation}
S=S_{({\rm B})} + S_{({\rm A})} \,, \label{0plus}
\end{equation}
where $S_{({\rm B})}$ is the action functional for the background field (without axion ingredients), and the second contribution $S_{({\rm A})}$ contains the axion field.
Why we can consider such a hierarchy? It is well known that the dark matter controls 23\% of the cosmic energy, but its distribution is quasi-uniform in the whole Universe volume (the corresponding averaged mass-density is about $\rho_{\rm DM} \propto 10^{-20}-10^{-24} {\rm g}/{{\rm cm}^3}$). As for the baryon matter, it is condensed in compact objects, and the corresponding mass density can reach the values of the order $\rho_{\rm B} \propto 10^{15}{\rm g}/{{\rm cm}^3}$. Clearly, the axionic effects in the dense objects can be considered in the given gravitational background formed by the baryonic mass and strong magnetic field. We start with the description of the regular background model.

The action functional for the background model has the form
\begin{equation}
S_{({\rm B})} {=} \int d^4 x \sqrt{{-}g} \left[\frac{R+2\Lambda}{2\kappa}
+ \frac{1}{4} F^{mn}F_{mn}  + \frac{1}{4}{\cal R}^{ikmn}F_{ik} F_{mn} \right] \,,
\label{actB}
\end{equation}
where $R$ is the Ricci scalar; $\Lambda$ is the cosmological constant;  $F_{ik}$ is the Maxwell tensor.  The term ${\cal R}^{ikmn}$ introduces the first nonminimal three-parameter susceptibility tensor
\begin{equation}
{\cal R}^{ikmn} =  q_1 R g^{ikmn} + q_2 \Re^{ikmn} + q_3 R^{ikmn}
\,, \label{sus1}
\end{equation}
\begin{equation}
g^{ikmn} \equiv \frac{1}{2}(g^{im}g^{kn} {-} g^{in}g^{km}) \,,
\label{rrr}
\end{equation}
\begin{equation}
\Re^{ikmn} \equiv \frac{1}{2} (R^{im}g^{kn} {-} R^{in}g^{km} {+}
R^{kn}g^{im} {-} R^{km}g^{in}) \,, \label{rrrr}
\end{equation}
where $R^{mn}$ is the Ricci tensor,  and $R^{ikmn}$ is the Riemann tensor.
The constants $q_1$, $q_2$ and $q_3$ are the nonminimal coupling parameters (we use the notations introduced in \cite{BL05}). In the first paper concerning this subject \cite{Prasanna} two of these parameters were chosen as vanishing ($q_1=q_2=0$). Drummond and Hathrell in \cite{DH} have presented these parameters as the result of calculations in the one-loop approximation of Quantum Electrodynamics in a curved spacetime; these parameters have the form $q_1 = -q$, $q_2=\frac{13}{5}q$, $q_3= -\frac{2}{5} q$, $q= \frac{\alpha \lambda^2_e}{36 \pi}$, where $\alpha$ is the fine structure constant, and $\lambda_{\rm e}$ is the Compton wavelength of the electron. Basic properties of the nonminimal field theory were discussed in many papers, one can use, e.g., \cite{Go,HO} and \cite{BL05} for details and references.

When the spacetime platform is static and spherically symmetric, it is convenient to use the metric
\begin{equation}
ds^2 = N(r)dt^2 - \frac{dr^2}{N(r)} - r^2 \left(d\theta^2 + \sin^2{\theta}{d\varphi}^2 \right)\,. \label{0metric}
\end{equation}
When three coupling parameters $q_1$, $q_2$ and $q_3$ are linked as follows:
\begin{equation}
q_1= -q  \,, \quad q_2 = 4q \,, \quad q_3 = -6q \,, \quad q >0 \,,
\label{0metricqqq}
\end{equation}
there exists the exact solution to the gravity field equations, for which (see \cite{R1,R2,R3,R4,R5})
\begin{equation}
N(r) = 1 + \frac{r^4}{r^4+2q Q^2}\left[-\frac{2M}{r}+ \frac{Q^2}{r^2} - \frac{\Lambda}{3}r^2 \right] \,.\label{N00}
\end{equation}
For the monopole-type magnetic field the Maxwell tensor has only one component
\begin{equation}
F_{\theta \varphi}= Q \sin{\theta} \,.
\label{0Fmn}
\end{equation}
Here and below $Q$ is the magnetic charge, $M$ is the asymptotic mass of the object.  The first invariant of the electromagnetic field $\frac12 F_{mn}F^{mn}$ is equal to $\frac{Q^2}{r^4}$; the second (pseudo)invariant $\frac12 F^{*}_{mn}F^{mn}$, where $F^{*}_{mn}$ is the tensor dual to the Maxwell one, is equal to zero.
These exact solutions behaves standardly in the asymptotic regime. The invariant $\frac12 F_{mn}F^{mn} = \frac{Q^2}{r^4}$ tends to zero, when $r \to \infty$. Far from the center the metric coefficient $N$ has the Minkowski asymptote $N \to 1-\frac{2M}{r}$, when $\Lambda=0$, the anti de Sitter asymptote $N \to  \frac{|\Lambda|}{3}r^2$, when $\Lambda < 0$, and possesses the cosmological horizon at $r=r_{\Lambda}$, where $N(r_{\Lambda})=0$, when $\Lambda>0$.

\subsubsection{Regularity at the center}

The interest to the exact solution (\ref{N00}) is connected with its behavior at the center of the object, $r=0$. Near the center one can use the decomposition
\begin{equation}
N(r\to 0) \approx 1+ \frac{r^2}{2q} - \frac{M}{q Q^2} r^3 - \frac{r^6}{2qQ^2}\left(\frac{\Lambda}{3}+ \frac{1}{2q} \right) \,,
\label{00N01}
\end{equation}
and we obtain the formulas, which will be necessary below in the work with the nonminimal susceptibility tensors.
As a reference, we provide, first, the formulas for the metric and its derivatives:
\begin{equation}
N(0)=1 \,, \quad N^{\prime}(0)=0 \,, \quad  N^{\prime \prime}(0)=\frac{1}{q} \,,
\label{spr1}
\end{equation}
second, the formulas for the components of the Riemann tensor:
$$
R^{0r}_{\ \ 0r} =\frac12 N^{\prime \prime} \ \ \Rightarrow   \frac{1}{2q}\,, \quad R^{\theta \varphi}_{\ \ \theta \varphi} = \frac{1}{r^2}(N-1) \ \ \Rightarrow  \frac{1}{2q} \,,
$$
\begin{equation}
R^{0 \theta}_{\ \ 0 \theta} = R^{0 \varphi}_{\ \ 0 \varphi} = R^{r \theta}_{\ \ r \theta} = R^{r \varphi}_{\ \ r \varphi} = \frac{1}{2r} N^{\prime} \ \ \Rightarrow   \frac{1}{2q} \,,
\label{spr2}
\end{equation}
third, the formulas for the components of the Ricci tensor
\begin{equation}
R^0_0 = R^r_r = \frac12 N^{\prime \prime} + \frac{1}{r} N^{\prime} \ \ \Rightarrow   \frac{3}{2q} \,, \quad R^{\theta}_{\theta}= R^{\varphi}_{\varphi}= \frac{1}{r} N^{\prime} + \frac{1}{r^2}(N-1) \ \ \Rightarrow   \frac{3}{2q} \,,
\label{spr3}
\end{equation}
and finally, the value of the Ricci scalar:
\begin{equation}
R = N^{\prime \prime}+ \frac{4}{r} N^{\prime}+ \frac{2}{r^2}(N-1) \ \ \Rightarrow   \frac{6}{q} \,.
\label{spr4}
\end{equation}
Thus, the quadratic curvature invariants: $R^2$, $R_{mn}R^{mn}$, $R_{ikmn} R^{ikmn}$ are finite at the center and are the constants proportional to $\frac{1}{q^2}$. In contrast, the first invariant of the electromagnetic field $\frac12 F_{mn}F^{mn}$ takes infinite value at the center.
The main question is the following: do the axion field and axionically induced electric field inherit the property of the gravity field, i.e., are regular, or they behave as the monopole - type magnetic field, i.e., become infinite at the center? We answer this question below, and show that the regular solutions for the axion and electric field exist for the special choice of the nonminimal coupling parameters.

\subsubsection{The model without event horizons}

Depending on the values of the parameters $q$, $M$, $|Q|$, $\Lambda$, the metric coefficient (\ref{N00}) can possess up to three horizons (see \cite{R5} for details). The third (the most distant) horizon appears as the cosmological one, when $\Lambda >0$; other two horizons are the inner and outer event horizons, which can, in principle, coincide for some critical values of the asymptotic mass. As it was shown in \cite{R5}, there are solutions, which have no event horizons; for the case $\Lambda >0$ these solutions are depicted as the curves I,II and III on the panel (a) of the Fig.2 in \cite{R5}; for the case $\Lambda=0$ the solutions of this type are presented as the lines I,II on the Fig.3; for the case $\Lambda <0$ such solutions are depicted as the curves I,II and III on the Fig.4 in \cite{R5}.
In this work, first, we assume that the asymptotic mass $M$ is chosen so that there are no event horizons in the background spacetime under consideration; second, we consider the background spacetime with regular center. We have to stress once again that these assumptions are valid in the specific nonminimal model only, and only two dimensionless parameters, $\frac{2q}{Q^2}$ and $\Lambda Q^2$, remain arbitrary guiding parameters.

\subsection{The extended model including the axion and electric fields}

\subsubsection{Action functional}

The contribution of the axion field can be describes by the second part of the sum (\ref{0plus}) given by
\begin{equation}
S_{({\rm A})} {=} \int d^4 x \sqrt{{-}g} \left\{\frac14 \phi \left[ F^{mn}F^{*}_{mn} + {\chi}^{ikmn}_{({\rm
A})} F_{ik} F^{*}_{mn} \right]+ \Psi^2_0 \left[\frac12  \left(V{-} g^{mn} \nabla_m \phi \nabla_n \phi \right)
 {-} \Re^{mn}_{({\rm A})} \nabla_m
\phi \nabla_n \phi {+} \eta_{({\rm A})} R  \phi^2 \right] \right\} \,.
\label{actnm}
\end{equation}
The symbol $\phi$ denotes the pseudoscalar field, associated with the axionic dark matter. The quantity ${\chi}^{ikmn}_{({\rm A})}$
given by
\begin{equation}
{\chi}^{ikmn}_{({\rm A})} = Q_1 R g^{ikmn} + Q_2\Re^{ikmn} + Q_3 R^{ikmn} \,, \label{sus2}
\end{equation}
is the second nonminimal susceptibility tensor, which describes the linear coupling
of the dual tensor $F^{*}_{mn}$ with the curvature; it was introduced in \cite{BWTN2010}. This object is constructed in analogy with the first susceptibility tensor ${\cal R}^{ikmn}$ (\ref{sus1}), and $Q_1$, $Q_2$, $Q_3$ are constants analogous to the parameters $q_1$, $q_2$, $q_3$.
The tensor $\Re^{mn}_{({\rm A})}$, given by
\begin{equation}
\Re^{mn}_{({\rm A})} \equiv \frac{1}{2} \eta_1
\left(F^{ml}R^{n}_{\ l} + F^{nl}R^{m}_{\ l} \right) + \eta_2 R g^{mn} + \eta_3 R^{mn} \label{sus3} \,,
\end{equation}
can be indicated as the third nonminimal susceptibility tensor associated with the coupling of the pseudoscalar field
with curvature. This term describes effects
analogous to the derivative coupling in the nonminimal
scalar field theory \cite{derc1,derc2,derc3}. As for the coupling constant $\eta_{({\rm A})}$, it is an analog of the
coupling constant $\xi$ in the nonminimal scalar field theory \cite{FaraR}. As usual, $V$ describes the potential of the pseudoscalar field; its simplest representation is $V=m^2_{A}\phi^2$, where $m_{A}$ is the axion mass. The parameter $\Psi_0$ is reciprocal to the constant of the axion-photon coupling $g_{A \gamma \gamma}$, $\Psi_0 = \frac{1}{g_{A \gamma \gamma}}$.

\subsubsection{Symmetry of the susceptibility tensors}

The third susceptibility tensor is explicitly symmetric, i.e.,
\begin{equation}
\Re^{mn}_{({\rm A})} = \Re^{nm}_{({\rm A})} \,.
\label{sus3sym}
\end{equation}
The tensors ${\cal R}^{ikmn}$ and  ${\chi}^{ikmn}_{({\rm A})}$,
defined by (\ref{sus1}) and (\ref{sus2}), are skew-symmetric with
respect to transposition of the indices in the pairs $ik$ and $mn$. Also, the following relations take place
\begin{equation}
{\cal R}^{ikmn}= {\cal R}^{mnik} \,, \quad {\chi}^{ikmn}_{({\rm
A})} = {\chi}^{mnik}_{({\rm A})}\,. \label{sus8}
\end{equation}
Our additional ansatz is that the second susceptibility tensor is symmetric with respect to the left and right dualization:
\begin{equation}
{}^{*}{\chi}^{ikmn}_{({\rm A})} = {\chi}^{*ikmn}_{({\rm A})}  \Leftarrow \Rightarrow {}^{*}{\chi}^{*ikmn}_{({\rm A})} = - {\chi}^{ikmn}_{({\rm A})}  \,.
\label{sus27}
\end{equation}
This requirement imposes a restriction on the constants $Q_1$, $Q_2$, $Q_3$; in order to obtain it, we decompose standardly the Riemann tensor using the Weyl tensor ${\cal C}^{ikmn}$, and rewrite the second susceptibility tensor as follows:
\begin{gather}
{\chi}^{ikmn}_{({\rm A})} = Q_3 {\cal C}^{ikmn} +
(Q_2+Q_3)\Re^{ikmn} +\left( Q_1-\frac{1}{3}Q_3\right) R g^{ikmn}
\,. \label{sus34}
\end{gather}
Since the double dualization provides
\begin{equation}
{}^{*}{g}^{*ikmn} = - g^{ikmn}  \,, \quad
{}^{*}{\Re}^{*ikmn} =  {\Re}^{ikmn} - R g^{ikmn}  \,, \quad
{}^{*}{C}^{*ikmn} = -C^{ikmn} \,,
\label{sus31}
\end{equation}
and the only $\Re^{ikmn}$ does not possesses this symmetry, we have to require that $Q_2{+}Q_3=0$ (see \cite{Q} for more detail). In other words, we use below the two-parameter second susceptibility tensor ${\chi}^{ikmn}_{({\rm A})}$.

\subsubsection{Nonminimal electrodynamic equations}

Electrodynamic equations can be obtained by the variation procedure applied to the total action functional
(\ref{0plus}) with (\ref{actB}) and (\ref{actnm}) with respect to the electromagnetic potential $A_i$; these master equations have the standard
form
\begin{equation}
\nabla_k  H^{ik} = I^i \,. \label{max2}
\end{equation}
The excitation tensor $H^{ik}$ is of the form
\begin{equation}
H^{ik} \equiv F^{ik} + {\cal R}^{ikmn} F_{mn} + \phi \left(
F^{*ik} + {\chi}^{ikmn}_{({\rm A})} F^{*}_{mn} \right) \,,
\label{inducnm}
\end{equation}
and the four-vector of the effective electric current
\begin{equation}
I^i \equiv \frac{1}{2}\eta_1 \Psi^2_0 \nabla_k \left[ \left(R^{km} \nabla^i
\phi - R^{im} \nabla^k \phi \right) \nabla_m \phi \right]
\label{current}
\end{equation}
satisfies, clearly, the conservation law $\nabla_i I^i = 0$.
As usual, the equation (\ref{max2}) has to be supplemented by the equation
\begin{equation}
\nabla_k F^{*ik}  =0 \,.
\label{8}
\end{equation}
When $\phi=0$, the equations (\ref{max2}) and (\ref{8}) give the background nonminimal electrodynamic equations, the solution to which coincides with (\ref{0Fmn}).

\subsubsection{Nonminimal master equation for the pseudoscalar field}

Nonminimally extended master equation for the pseudoscalar $\phi$
can be obtained by variation of the action functional with respect to $\phi$, yielding
\begin{equation}
\nabla_m \left[ \left( g^{mn} + \Re^{mn}_{({\rm A})} \right)
\nabla_n \phi \right] + \left[m^2_{({\rm A })} + \eta_{({\rm A})}
R \right] \phi = - \frac{1}{4 \Psi^2_0} \left[F^{mn}F^{*}_{mn} +
{\chi}^{ikmn}_{({\rm A})} \ F_{ik} F^{*}_{mn}\right] \,. \label{A0}
\end{equation}
The electromagnetic source in the right-hand side of this equation contains minimal and nonminimal contributions.

\subsection{Reduced master equations of the nonminimal axion magnetoelectrostatics}

\subsubsection{Reduced electrostatic equations}

We assume that the pseudoscalar and electromagnetic  fields inherit the spacetime symmetry. This means that $\phi$ is the function of the radial variable only, $\phi(r)$. As for the potential of the electromagnetic field, it can be presented in the form
\begin{equation}
A_i = \delta_i^0 A_0(r) + \delta_i^{\varphi} A_{\varphi}(\theta).
\label{poten}
\end{equation}
First of all, we consider the reduced electric current (\ref{current})
\begin{equation}
I^i \equiv \frac{1}{2r^2}\eta_1 \Psi^2_0 \frac{d}{dr} \left[ r^2 \left(R^{rr} g^{ir}-R^{ir}g^{rr}\right) \left(\frac{d\phi}{dr}\right)^2 \right]  \,.
\label{2current}
\end{equation}
Clearly, it vanishes in the fields with given symmetry.
Due to the model symmetry, only the  $H^{ir}$ and $H^{i\theta}$ components of the induction tensor enter the reduced electrodynamic equations (\ref{max2}):
\begin{equation}
\frac{1}{r^2} \partial_r \left(r^2 H^{ir} \right) + \frac{1}{\sin{\theta}} \partial_{\theta} \left(\sin{\theta} H^{i\theta} \right) =0 \,,
\label{mx5}
\end{equation}
where the explicit expressions for these components have the form
\begin{equation}
H^{ir} = \delta^i_0 \left[\frac{dA_0}{dr}\left(1+2{\cal R}^{0r}_{\ \ 0r} \right) + \frac{Q \phi}{r^2}\left(1+2\chi^{0r}_{\ \ 0r} \right) \right] \,,
\label{mx6}
\end{equation}
\begin{equation}
H^{i\theta} = \frac{1}{\sin{\theta}}\delta^i_{\varphi} \left[-\frac{Q}{r^4}\left(1+2{\cal R}^{0r}_{\ \ 0r} \right) + \frac{\phi}{r^2} \frac{dA_0}{dr}\left(1+2\chi^{0r}_{\ \ 0r} \right) \right] \,.
\label{mx7}
\end{equation}
Direct calculations show that the component $A_{\varphi}$ again is of the form $A_{\varphi}=Q(1-\cos{\theta})$, and the nonminimal electrodynamic equations can be now reduced to one equation  for the component $A_0(r)$:
\begin{equation}
\frac{d}{dr}\left[r^2 \frac{dA_0}{dr}\left(1+2{\cal R}^{0r}_{\ \ 0r} \right) + Q \phi \left(1+2\chi^{0r}_{\ \ 0r} \right)\right] = 0 \,.
\label{mx9}
\end{equation}
The solution to this equation contains the electric field $E(r) \equiv F_{r0}  = \frac{dA_0}{dr}$, and this solution written as
\begin{equation}
\frac{dA_0}{dr}\left(1+2{\cal R}^{0r}_{\ \ 0r} \right) = \frac{1}{r^2} \left[K - Q \phi \left(1+2\chi^{0r}_{\ \ 0r} \right) \right]
\label{mx8}
\end{equation}
contains the integration constant $K$, which can be standardly interpreted as an electric charge registered by the distant observer.
Direct calculation of the necessary components of the first and second susceptibilities yields
\begin{equation}
{\cal R}^{0r}_{\ \ 0r}= q\left[-\frac32N^{\prime \prime} + \frac{2}{r} N^{\prime} - \frac{1}{r^2}(N-1)   \right] \,,
\label{sus98}
\end{equation}
\begin{equation}
\chi^{0r}_{\ \ 0r} = \chi^{\theta \varphi}_{\ \ \theta \varphi} = Q_1\left[\frac12 N^{\prime \prime} + \frac{2}{r} N^{\prime} + \frac{1}{r^2}(N-1) \right] -  \frac{1}{r}Q_3 N^{\prime} \,.
\label{sus99}
\end{equation}
Here we took into account that $Q_2=-Q_3$.

\subsubsection{Reduced equation for the axion field}

The equation (\ref{A0}) takes now the form
\begin{equation}
\frac{1}{r^2}\frac{d}{dr} \left\{r^2 N \left(\frac{d \phi}{dr}\right)\left[1+\eta_2 R + \eta_3 R^r_r \right]\right\} - \left(m_{A}^{2} + \eta_A R \right)\phi  + \frac{Q}{r^2\Psi^2_0} \left(\frac{dA_0}{dr}\right) \left(1+ 2\chi^{0r}_{\ \ 0r} \right) = 0 \,.
\label{A2}
\end{equation}
If we take the expression for the electric field $\frac{dA_0}{dr}$ from (\ref{mx8}) and put it into (\ref{A2}), we obtain the key equation for the pseudoscalar (axion) field:
\begin{equation}
\frac{1}{r^2}\frac{d}{dr} \left\{r^2 N \left(\frac{d \phi}{dr}\right)\left[1+\eta_2 R + \eta_3 R^r_r \right]\right\} - \left(m_{A}^{2} + \eta_A R \right)\phi +  \frac{Q \left(1+ 2\chi^{0r}_{\ \ 0r} \right)}{r^4\Psi^2_0\left(1+2{\cal R}^{0r}_{\ \ 0r} \right)}\left[K- Q\phi \left(1+2\chi^{0r}_{\ \ 0r} \right)  \right] = 0 \,,
\label{A4}
\end{equation}
where all the auxiliary functions $N(r)$, $R$, $R^r_r$, ${\cal R}^{0r}_{\ \ 0r}$, $\chi^{0r}_{\ \ 0r}$ are already calculated and presented above.

\section{Solutions regular at the center}

We search for the solutions to the key equations (\ref{mx8}) and (\ref{A4}), which are regular at the center $r=0$, thus inheriting the regularity of the gravitational field. Clearly, it is possible for a special choice of the coupling parameters $Q_1$ and $Q_3$ and of the integration constant $K$.

\subsection{The model with $Q_1=\frac32 q$, $Q_2= -5q$, $Q_3 = 5q$ and $K=0$}

When $K=0$, the whole electric charge of the object is equal to zero and thus the Coulombian part of the electric field $\frac{K}{r^2}$ is absent.
When we put $Q_1=\frac32 q$ and $Q_3 = 5q$ into the formulas (\ref{sus98}) and (\ref{sus99}), we obtain that for small values of the radial variable $r \to 0$
\begin{equation}
1+2 {\cal R}^{0r}_{\ \ 0r} \approx 1+ \frac{8Mr}{Q^2} \,, \quad 1+ 2\chi^{0r}_{\ \ 0r} \approx - \frac{6r^4}{Q^2}\left(\frac{\Lambda}{3}+ \frac{1}{2q} \right)  \,.
\label{A44}
\end{equation}
Then the key equation for the axion field in the near zone, rewritten in the leading order approximation, converts into
\begin{equation}
\frac{1}{r^2}\frac{d}{dr} \left(r^2  \frac{d \phi}{dr}\right) - \mu^2_0 \phi  = 0 \,,
\label{A55}
\end{equation}
where the auxiliary parameter $\mu_0$ plays the role  of an effective mass of the axion nonminimally coupled to the curvature.
\begin{equation}
\mu_0 \equiv \sqrt{\frac{m_{A}^{2} + \frac{6\eta_A}{q}}{1+\frac{3}{2q}(4\eta_2 + \eta_3)}} \,.
\label{A551}
\end{equation}
The solution to (\ref{A55}), regular at the center, is well known
\begin{equation}
\phi(r)= \phi(0)\frac{{\rm sh}\mu_0 r}{\mu_0 r} \,.
\label{A552}
\end{equation}
In fact, in the near zone we have to use the approximation
\begin{equation}
\phi(r) \approx \phi(0)\left(1 + \frac{\mu^2_0 r^2}{6}\right)\,,
\label{A553}
\end{equation}
for which $\phi(0) \neq \infty$ and $\phi^{\prime}(0) = 0$. Similarly, we obtain from (\ref{mx8}) the electric field
\begin{equation}
E(r) = \frac{dA_0}{dr} \approx  \phi(0) \ \frac{6r^2}{Q}\left(\frac{\Lambda}{3}+ \frac{1}{2q} \right) \,.
\label{A554}
\end{equation}
We see that $E(0) = 0$ and $E^{\prime}(0) = 0$, i.e., the electric field is also regular at the center. Behavior of the pseudoscalar and electric fields in the full interval $0<r<\infty$ can be analyzed only numerically. For the illustration of this behavior in the next subsection we consider the particular regular nonminimal model with $\Lambda=0$ and $M=0$.

\subsection{Numerical analysis of the particular model with $\Lambda=0$, $M=0$}

For the analysis of the indicated particular model we have to rewrite the key equations using the following exact formulas:
\begin{equation}
N= \frac{r^4 + r^2 Q^2 +2q Q^2}{r^4+2q Q^2} \,,
\label{vN00}
\end{equation}
\begin{equation}
\frac{N-1}{r^2}= \frac{Q^2}{2qQ^2+r^4} \,, \quad \frac{N^{\prime}}{r} = 2Q^2 \frac{(2qQ^2-r^4)}{(2qQ^2+r^4)^2} \,, \quad N^{\prime \prime} = \frac{2Q^2}{(2qQ^2+r^4)^3}(3r^8-24qQ^2 r^4 + 4 q^2 Q^4) \,.
\label{vN005}
\end{equation}
Now the key equation for the axion field can be rewritten in the form
\begin{equation}
\phi^{\prime \prime} + \frac{\Pi^{\prime}}{\Pi} \phi^{\prime} - B(r) \phi =0\,,
\label{key33}
\end{equation}
where we introduced the coefficients $\Pi(r)$ and $B(r)$ as follows:
\begin{equation}
\Pi(r) \equiv r^2 N \left(1+\eta_2 R + \eta_3 R^r_r \right) \,,
\label{key34}
\end{equation}
\begin{equation}
B(r) \equiv \frac{r^4\Psi^2_0 \left(m_{A}^{2} + \eta_A R\right)\left(1+2{\cal R}^{0r}_{\ \ 0r} \right) +  Q^2 \left(1+2\chi^{0r}_{\ \ 0r} \right)^2}{N r^4 \Psi^2_0\left(1+2{\cal R}^{0r}_{\ \ 0r} \right)\left(1+\eta_2 R + \eta_3 R^r_r \right)} \,.
\label{key35}
\end{equation}
To facilitate the numerical analysis we use the replacement $r=|Q|x$. With this replacement we obtain the dimensionless quantities, and two guiding parameters: $\xi = \frac{2q}{Q^2}$ and $\eta = \frac{\eta_A}{Q^2}$.
For the facilitated understanding, we have to stress, that, first, the term
\begin{equation}
\left(m_{A}^{2} + \eta_A R\right) = m_{A}^{2} +  \frac{8\eta_A q Q^4(6qQ^2-5r^4)}{(2qQ^2+r^4)^3} = m_{A}^{2} +  \frac{4\eta_A  \xi(3\xi-5x^4)}{(\xi+x^4)^3}
\label{key36}
\end{equation}
is regular; at the center it takes the value
$m_{A}^{2} + \eta_A \frac{6}{q}$; second, the term
\begin{equation}
\left(1+2{\cal R}^{0r}_{\ \ 0r} \right) = \frac{x^{12}-11 \xi x^8 +37 \xi^2 x^4 + \xi^3}{(\xi+x^4)^3}
\label{key37}
\end{equation}
is also regular, has no real positive roots, and takes the value equal to $1$ at the center; third, the term
\begin{equation}
\left(1+2\chi^{0r}_{\ \ 0r} \right) = r^4\frac{(r^8 + 26qQ^2 r^4-48q^2Q^4)}{(2qQ^2+r^4)^3} = x^4\frac{(x^8 + 13\xi x^4-12 \xi^2)}{(\xi+x^4)^3}
\label{key39}
\end{equation}
vanishes at the center. Concerning the last ingredient of the formula (\ref{key35})
\begin{equation}
\left(1+\eta_2 R + \eta_3 R^r_r \right) \Rightarrow 1+ N^{\prime \prime} \left(\eta_2+\frac12 \eta_3 \right) + \frac{N^{\prime}}{r}(4\eta_2+ \eta_3) + \frac{2}{r^2}(N-1) \eta_2 \,,
\label{key391}
\end{equation}
for an example of the numerical simulation we use the particular relationships $\eta_3=-2\eta_2$, $q_2=q$, yielding
\begin{equation}
\left(1+\eta_2 R + \eta_3 R^r_r \right) = \frac{r^8 + 2qQ^2 r^4 + 16q^2Q^4}{(2qQ^2+r^4)^2} = \frac{x^8 + \xi x^4 + 4\xi^2}{(\xi+x^4)^2}\,.
\label{key392}
\end{equation}
This quantity is also regular, has no real positive roots, and is equal to 4 at the center. Thus, the coefficient $B(r)$ in (\ref{key33}) is regular at $0<r<\infty$.

Keeping in mind these properties, and based on the structure of the coefficients (\ref{key34}) and (\ref{key35}) in the key equation for the axion field (\ref{key33}), as well as in the key equation for the electric field (\ref{mx8}), we made the numerical simulation of the corresponding solutions by varying the dimensionless guiding parameter $\xi=\frac{2q}{Q^2}$. The results are illustrated on the Fig.1, Fig.2, Fig.3. On the Fig.1 we depicted the profiles of the metric coefficient $N(x,\xi)$; on the Fig.2 the typical profiles of the axion field are illustrated; on the Fig.3 we presented the typical profiles of the axionically induced electric field.
\begin{figure}[t]
	\includegraphics[width=130mm,height=90mm]{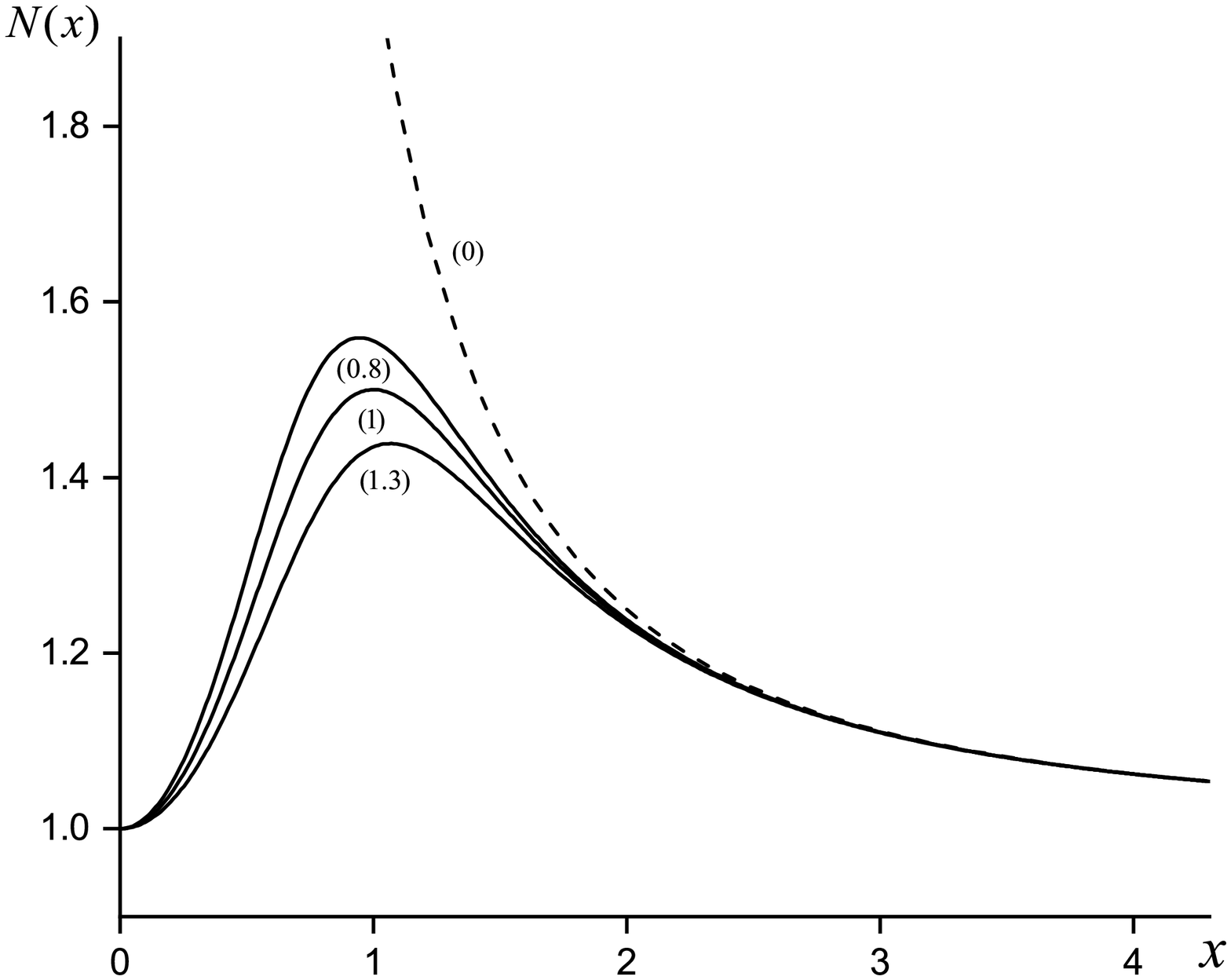}
	\caption{Illustration of the profiles of the metric coefficient $N(x,\xi)=1+\frac{x^2}{x^4+\xi}$, which is obtained from (\ref{N00}), when $\Lambda=0$ and $M=0$. The dimensionless radial variable $x$ is introduced as $r=x |Q|$, and the guiding parameter $\xi = \frac{2q}{Q^2}$ is indicated in the parentheses near the curve. All the curves start with $N(0)=1$ and tend asymptotically to $N(\infty)=1$. The function $N(x,\xi)$ is regular and has no zeros (i.e., horizons are absent). The dotted line relates to the case $\xi=0$ and illustrates the special solution of the Reissner-Nordstr\"om type singular at the center.}
\end{figure}
\begin{figure}[t]
	\includegraphics[width=130mm,height=90mm]{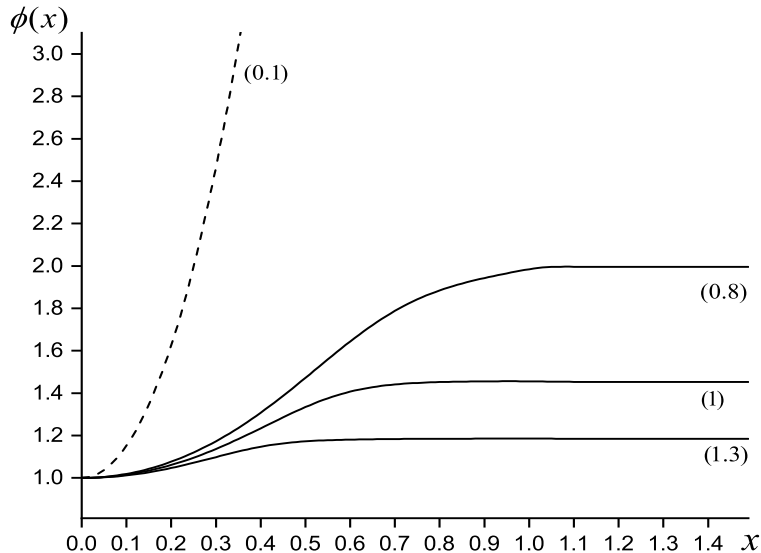}
	\caption{Illustration of the profiles of the axion field, presented by the solutions to the equation (\ref{A4}) with $Q_1=\frac32 q$, $Q_2= -5q$, $Q_3 = 5q$ and $K=0$, on the regular spacetime background with $\Lambda=0$, $M=0$. The solutions $\phi(x)$ are regular at the center, and tend asymptotically to constants; for the illustration we took $\phi(0)=1$ for all the solutions and obtain, respectively, different asymptotic values $\phi(\infty)$. When we study the solutions with $\phi(\infty)$ coinciding for all the curves, we obtain the similar curves, however, the starting values $\phi(0)$ become different and finite. The dotted line illustrates the fact that for small values of the parameter $\xi=\frac{2q}{Q^2}$, indicated in parentheses near the curve, the ratio $\frac{\phi(\infty)}{\phi(0)}$ grows significantly.}
\end{figure}
\begin{figure}[t]
	\includegraphics[width=130mm,height=90mm]{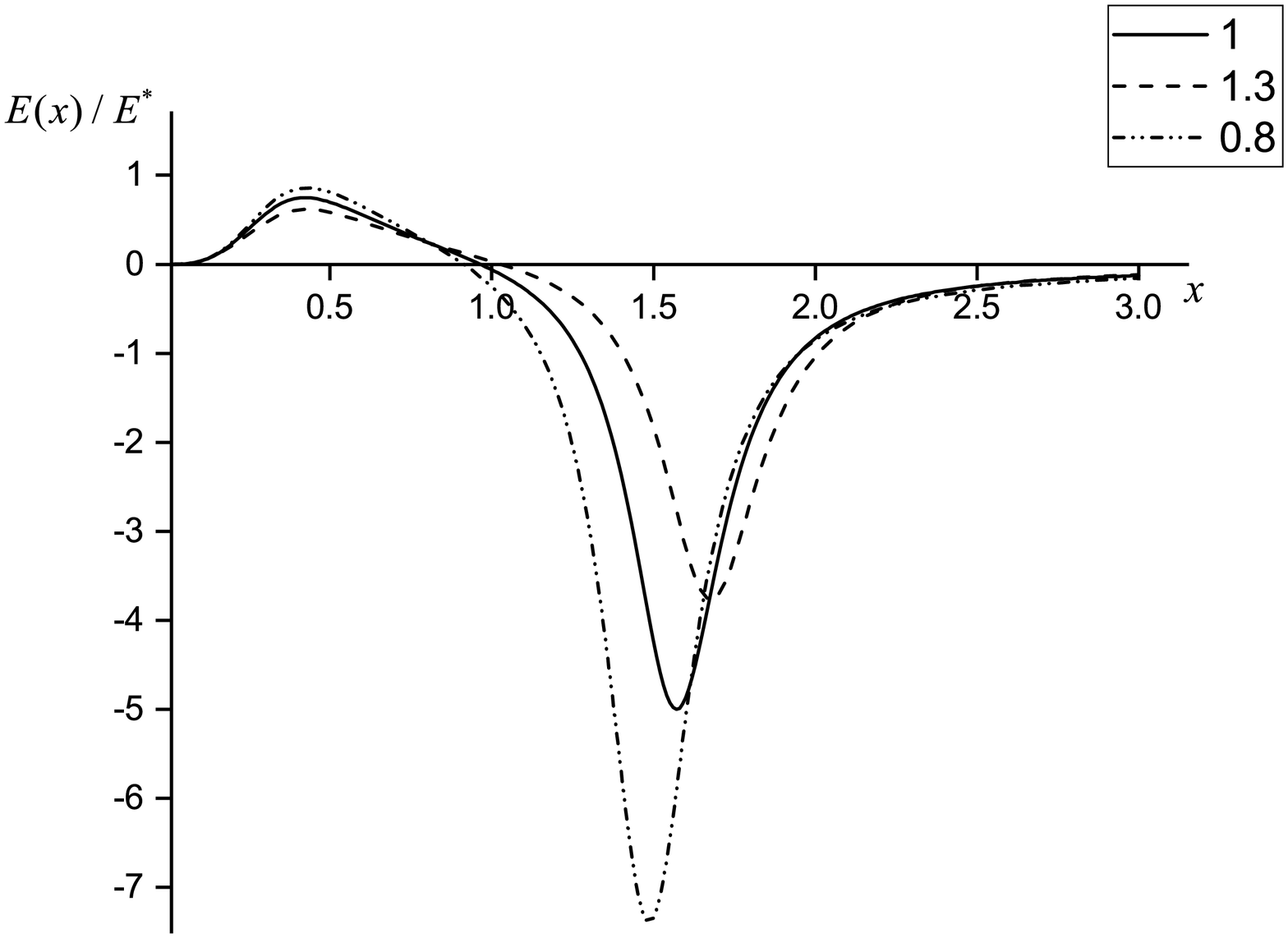}
	\caption{Illustration of the profiles of the dimensionless values of the axionically induced electric field, $E(x)/E^{*}$, where $E^{*}=1/Q$ (in the accepted units). All the curves, presenting the electric field, start from zero value at the center of the dyon, and tend to zero asymptotically far from the center. Each solution has one zero at $x \neq 0$; i.e., the induced electric field changes the sign once, possesses one potential hole and behaves as the electric field in a simple layer. When the positive guiding parameter $\xi = \frac{2q}{Q^2}$ decreases, the potential hole deepens, and the internal barrier, associated with the maximum of the electric field, grows. When $\xi = 0$, the electric field at the center becomes infinite ($E(0)= - \infty$). }
\end{figure}

\section{Discussion}

The radial electric field emerging near the magnetic monopole, surrounded by the dark matter halo, can be considered as a marker, which indicates that the dark matter consists of axions, or, at least, contains an axionic fraction. Observation of such an axionically induced electric field would be the indirect solution of the problem of identification of the particles, which form the dark matter. Since not only the axion-photon coupling can produce the electric field, we need to predict the existence of a distinctive feature, which is the undeniable attribute of the axionic influence. In the series of papers \cite{BG1,BG2,BG3,BG4} we elaborated the idea, that the field configuration, typical for the electric layer, in the near zone of the axionic dyon could be such a distinctive feature. Indeed, if there is a domain in the dyon vicinity, in which the electric field changes the sign, one can try to find the corresponding imprints in the object spectrum. Elaborating this idea we studied, first of all, the profiles of the axion, electric and distorted magnetic fields in the outer zone of the dyon, i.e., at $r>R>r_{\rm outer}$, where $R$ is the radius of the solid body of the magnetic star, and $r_{\rm outer}$ is the radius of the outer horizon of the object (see \cite{BG1,BG2}). The analysis of the stepwise distribution of the axion field near the black holes \cite{BG3} was the next step in this investigation. Then we studied the fold-like structures near the axionic dyons without horizons but with the central (naked) singularity \cite{BG4}. Keeping in mind the idea of Penrose about the Cosmic Censorship \cite{Penrose}, and based on the results of the work \cite{Censor}, we decided that the nonminimal coupling can play the role of such Censor, and have analyzed in this paper the model with nonminimally coupled gravitational, axion and electric fields regular at the center of the axionic dyon. In other words, the presented paper gives us the final detail additional to the series of works \cite{BG1,BG2,BG3,BG4}. What are the main features of this analysis?

1. Even if the spacetime metric and curvature invariants are regular at the center of a nonminimal magnetic monopole (see Subsubsection IIA2), there is no guarantee, in general case, that the electric and axion fields also are regular. As we have shown (see Subsection IIIA), the regularity of the axion and electric fields can be provided, if and only if  we consider the additional nonminimal axion-photon coupling with the specific choice of the interaction constants: $Q_1=\frac32 q$, $Q_2= -5q$, $Q_3 = 5q$, where the coupling constants $Q_1$, $Q_2$, $Q_3$ form the second susceptibility tensor (\ref{sus2}), and $q$ is the coupling constant, which enters the first susceptibility tensor (see (\ref{sus1}) with (\ref{0metricqqq})). The bonus of this choice is that the regular electric field takes zero value at the center, $E(0)=0$, its first derivative vanishes, $E^{\prime}(0)=0$, and this point relates to the local minimum.

2. In all models of the axionic dyons (minimal and nonminimal) the asymptotic behavior of the pseudoscalar field is similar: the axion field $\phi(r)$ tends to a constant, $\phi(\infty)$, and its derivative $\phi^{\prime}(r)$ tends to zero. The main difference reveals in the zone near the center. Minimal models (see, e.g., \cite{BG4}) predict the infinite value at the center, $\phi(0)=\infty$, while the model of nonminimal axion-photon coupling admits the existence of the regular axion field with a finite value at the center, $\phi(0)$. As it was illustrated by the Fig.2, the ratio $\frac{\phi(\infty)}{\phi(0)}$ depends on the guiding parameter $\xi=\frac{2q}{Q^2}$ and can be found numerically. This means that, if we fix the value $\phi(\infty)$ and connect it with the dark matter parameters far from the dyon, we can predict the value $\phi(0)$ at the center.

3. If we consider the minimal models of the axionic dyons, the axionically induced electric field $E(r)$ is singular at the center (see, e.g., \cite{BG4}). In the framework of the nonminimal model we obtain that $E(0)=0$ and $E^{\prime}(0)=0$. Since the axionically induced electric field can not be trivial, and asymptotic value $E(\infty)$ also is equal to zero, we conclude that there are extrema on the profiles of the electric field. Indeed, as it was shown by numerical analysis (see Fig.3.), there are the maximum, zero point and minimum on the profiles of the electric field for all values of the guiding parameter $\xi$. Clearly, the electric field changes the sign once, i.e., in the near zone there exists a field configuration similar to the one in a simple electric layer. The depth of the minimum and height of the maximum depend on the value of the guiding parameter $\xi=\frac{2q}{Q^2}$; when $\xi$ decreases, the electric potential hole deepens.

The presented nonminimal model of the axionic dyon needs numerical estimations and observational testing, however, this activity is out of the framework of this paper.

\begin{acknowledgments}
The work was supported by Russian Foundation for Basic Research (Project No. 20-52-05009), and, partially, by the Program of Competitive Growth of Kazan Federal University.
\end{acknowledgments}

\end{document}